# High-Mobility and High-Optical Quality Atomically Thin WS$_2$

Francesco Reale[a], Pawel Palczynski[a], Iddo Amit[b], Gareth F. Jones[b], Jake D. Mehew[b], Agnes Bacon[b], Na Ni[a], Peter C. Sherrell[a], Stefano Agnoli[c], Monica F. Craciun[d], Saverio Russo[b], Cecilia Mattevi*[a]

[a] Department of Materials, Imperial College London, SW7 2AZ, UK

[b] Centre for Graphene Science, Department of Physics, University of Exeter, Stocker Road, Exeter, EX4 4QL, UK

[c] Department of Chemical Sciences, University of Padua, Via F. Marzolo 1, 35131 Padua, Italy

[d] Centre for Graphene Science, Department of Engineering, University of Exeter, North Park Road, Exeter, EX4 4QF, UK

*c.mattevi@imperial.ac.uk

ABSTRACT

The rise of atomically thin materials has the potential to enable a paradigm shift in modern technologies by introducing multi-functional materials in the semiconductor industry. To date the growth of high quality atomically thin semiconductors (e.g. WS$_2$) is one of the most pressing challenges to unleash the potential of these materials and the growth of mono- or bi-layers with high crystal quality is yet to see its full realization. Here, we show that the novel use of molecular precursors in the controlled synthesis of mono- and bi-layer WS$_2$ leads to superior material quality compared to the widely used topotactic transformation of WO$_3$-based precursors. Record high room temperature charge carrier mobility up to 52 cm$^2$/Vs and ultra-sharp photoluminescence linewidth of just 36 meV over submillimeter areas demonstrate that the quality of this material supersedes also that of naturally occurring materials. By exploiting surface diffusion kinetics of W and S species adsorbed onto a substrate, a deterministic layer thickness control has also been achieved promoting the design of scalable synthesis routes.

INTRODUCTION

Atomically thin layers of metal group VI disulfides and diselenides ($MoS_2$, $WS_2$, $WSe_2$, $MoSe_2$,) are being extensively investigated as they present unconventional optoelectronic properties compared to commonly use low-dimensional semiconductors[1,2]. In the bulk form they are layered compounds formed of covalently bonded chalcogen and metal atoms forming tri-atomic layers, which are held together by van der Waals forces[3]. An individual tri-atom thick layer presents a direct band gap in the visible-near IR range[4] on the contrary to the bulk, which manifests an indirect electronic band gap. Monolayer sulphides and selenides show strong light absorption from the visible to the near IR range[1,5,6], valley polarization[7,8], second-harmonic generation[9], tightly bound excitons[10] and strong spin-orbit interaction[11,12]. These properties arise from their intrinsic two-dimensional nature inherently free from dangling bonds and their particular d-orbitals configuration[3,13]. Further, given their atomically thin nature they are mechanically flexible and they can sustain tensile strain of 20%[14,15].

One of the most promising transition metal dichalcogenides (TMDCs) is $WS_2$ owing to light emission in the monolayer form at ~ 2 eV and the low level of toxicity of growth processes. Any envisioned application relies on materials with high crystal and optical quality extended over wafer-size areas. Chemical vapour deposition (CVD) is a scalable method for materials synthesis and it is being widely employed for TMDCs[16,17]. The synthesis of tungsten-based materials has revealed to be challenging and generally leading to isolated flakes of lateral size between 5-40 µm[4,18-23]. Monolayer $WS_2$ films extended over centimeter-sized areas has been demonstrated[24,25], however with compromised crystal quality as indicated by the low carrier mobilities. The growth is typically performed by co-evaporating sulfur powder and a W-precursor in a horizontal tubular furnace in presence of a carrier gas. Until now the synthesis of $WS_2$ has been achieved predominantly using $WO_3$ and S powders[4,18-23] at temperatures greater than 900 °C. This involves a topotactic transformation[26] which normally yields to sparsely distributed $WS_2$ domains onto an amorphous[4,18,20,22,23] or crystalline substrate[19,21,27].

Here we demonstrate the synthesis of high quality monolayer $WS_2$ using carbon-free molecular precursors. The high crystal quality is manifested by the record high charge carrier mobilities of mono- and bi-layer $WS_2$ and ultra-sharp PL linewidth at room temperature, which are superior to those of naturally occurring materials. The growth is enabled by molecular precursors, which lead to a complete sulfidization of W and formation of $WS_2$ with lower number of defects compared to the traditionally used topotactic conversion of $WO_3$.

**RESULTS AND DISCUSSION**

The synthesis of $WS_2$ was performed starting from commercial powders of either $H_2WO_4$ (hydrated tungsten oxide), sulfur and where indicated, we have introduced NaCl. W and S precursors were placed in two separate crucibles well-spaced in a quartz tubular furnace (Supporting Information, Figure S1a) and heated up independently using different controllers as reported in Figure S1b. The heating profile of S powders has been optimized to ensure maximum supply when the W-precursors start evaporating. WS$_2$ was grown on Si/ SiO$_2$ (285 nm) substrates loaded in the downstream zone of the tubular furnace. Altering the chemistry of decomposition of tungsten oxide species, we could achieve synthesis of monolayer WS$_2$ over larger area coverage, at low temperatures and with low amount of defects.

Optical micrographs of WS$_2$ monolayers (Figure 1) grown using different tungsten oxide precursors systems at different temperatures (950 °C, 850 °C and 750 °C) show distinctively increasing lateral size of the triangles and increasingly facilitated synthesis at low temperatures from WO$_3$ to WO$_3$-NaCl and H$_2$WO$_4$-NaCl. The possibility to grow WS$_2$ from WO$_3$ at temperatures not lower than 950 °C is notoriously attributed to the high sublimation temperature of the oxide[17,28-30]. In specific, the size of the WS$_2$ triangles increases from ~10 μm to 60 μm to 200 μm and showing also areas of continuous polycrystalline monolayered coverage of ~ 0.8 mm (Figure S2). Further, the WS$_2$ synthesis at temperatures as low as 750 °C was enabled only by the precursors system of H$_2$WO$_4$-NaCl with remarkable triangular size of ~100 μm. In addition, increasing the growth pressure (from ~1 mbar to 13 mbar) at 950 °C, bilayered WS$_2$ flakes are preferentially formed (Figure S3) using the H$_2$WO$_4$-NaCl system.

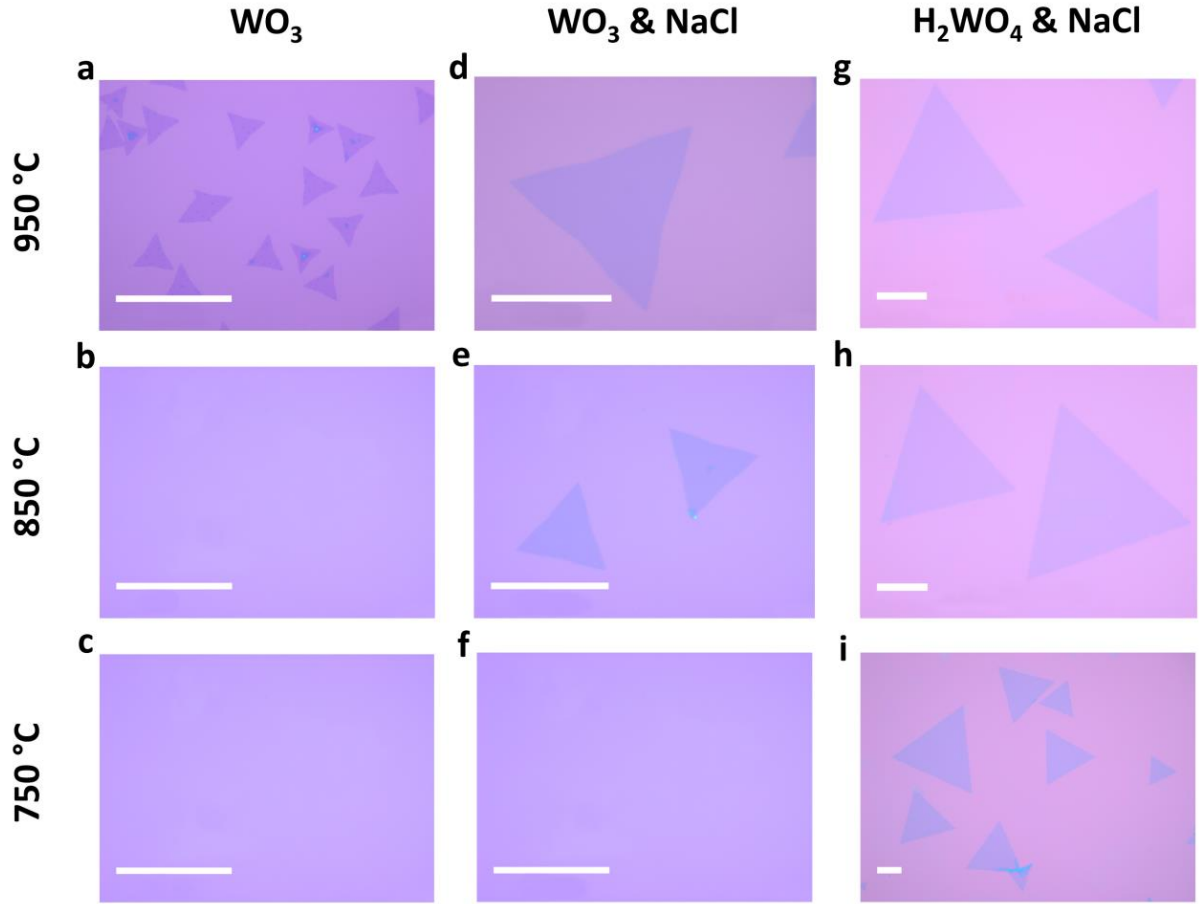

*Figure 1. Optical micrographs of WS$_2$ triangles grown on SiO$_2$/Si substrates at different temperatures and using different precursors: (a) WO$_3$ at 950 °C; (b) WO$_3$ at 850 °C; (c) WO$_3$ at 750 °C; (d) WO$_3$ + NaCl at 950 °C; (e) WO$_3$ + NaCl at 850 °C; (f) WO$_3$ + NaCl at 750 °C; (g) H$_2$WO$_4$ + NaCl at 950 °C; (h) H$_2$WO$_4$ + NaCl at 850 °C; (i) H$_2$WO$_4$ + NaCl at 750 °C. Scale bar is 50 μm.*

It is worth noting that larger WS$_2$ domains obtained at 950 °C as compared to 850 and 750 °C, can be explained in the light of the Robinson & Robin model[31]. At high temperature the diffusivity of the adsorbed precursors on the SiO$_2$ surface is favourable leading to the expansion of the existing domains. At the same time, desorption of absorbed species is higher than at lower temperatures, limiting the achievement of a supersaturation concentration and thus reducing the nucleation density.

From structural investigation of the reaction products in the different precursors systems we could elucidate the role played by the water intercalated in the $H_2WO_4$ and how this can enable the favorable synthesis at low temperature and with low density of defects as compared to the $WO_3$-based precursors. From X-ray diffraction (XRD) characterization (Supporting Information) it was possible to observe that the main products of the reactions between NaCl and either $H_2WO_4$ or $WO_3$ are similar: $Na_xW_yO_z$ and tungsten oxychlorides ($WClO_4$ and $WO_2Cl_2$). However the reactions occur at significantly lower temperatures for $H_2WO_4$ compared to $WO_3$. While $Na_xW_yO_z$ species possesses a high evaporation temperature, as they are found in the crucible (Figure S5) at the end of the synthesis of $WS_2$, the tungsten oxychlorides are volatile (Figure S4, S5) and they can possibly play a key role in promoting the synthesis of $WS_2$. Indeed the system $H_2WO_4$-NaCl is likely to enable the formation of tungsten-oxyhalide species (i.e., $WO_2Cl_2$, $WOCl_4$) at lower temperatures than $WO_3$-NaCl, as NaCl dissociation is promoted by the $H_2O$ molecules gradually released by $H_2WO_4$ upon heating. On the bases of previous studies on the synthesis of $WS_2$ bulk crystals, the formation of tungsten oxychlorides ($WO_2Cl_2$ and $WOCl_4$) is indeed likely to occur with higher chances with respect to the formation of metal halides (e.g. $WCl_6$)[32,33]. Tungsten oxychlorides ($WO_2Cl_2$ and $WOCl_4$) can be volatile from 200 °C[34] and they can be sulfidized in vapour phase forming a few-atom clusters of W-S which can deposit onto the target substrate as adatom species[35] where they can form $WS_2$ via a diffusion-desorption mediated mechanism of nucleation and growth. $WOCl_4$ has been previously used[36] as precursor for the synthesis of $WS_2$ bulk films. Despite its strong tungsten-oxygen double bonds, $WOCl_4$ proved to be an effective precursor with a clean decomposition pathway without formation of tungsten oxysulfides. We have verified that using this precursor is indeed possible to obtain $WS_2$ at temperatures as low as 550 °C (Figure S6a). The key role played by the oxyhalide species becomes apparent if we try to grow $WS_2$ by using $H_2WO_4$ as single precursor. As this decomposes to form $WO_3$, only small $WS_2$ domains are observed with PL characteristics similar to the $WO_3$ precursor-led growth (Figure S6b).

High-resolution transmission electron microscopy (HRTEM) imaging confirms the high crystalline nature of the material (Figure 2a). The measured lattice constant is ~0.3 nm consistent with that of $2H-WS_2$ (a=0.318 nm). The Raman spectra of $WS_2$ obtained using the different precursor systems are shown in Figure 2b. All of the spectra exhibit two characteristic peaks located at ~(351±0.53) cm$^{-1}$ and ~(417.6±1) cm$^{-1}$, which can be attributed to 2LA-$E^1_{2g}$ and $A_{1g}$ Raman modes of pristine $WS_2$ monolayer[37,38]. Interestingly, the distribution of the peak positions is increasingly narrower from $WO_3$ precursor, $WO_3$+NaCl and to $H_2WO_4$+NaCl (Figure S8, S10). The Raman peak intensities are uniform across the entire triangle area (Figure S7, S9) and the frequency difference (Δν) between 2LA(M) and $A_{1g}$ modes is ~(66.5±0.53) cm$^{-1}$ (Figure S11), as expected for monolayer $WS_2$[37]. The AFM thickness profile analysis of $WS_2$ triangles confirms the monolayer (Figure 2c,d) and bilayer (Figure 2e,f) nature of the flakes, showing an edge step height of ~0.8 nm (Figure 2d) and ~1.6 nm (Figure 2f), respectively[39,40].

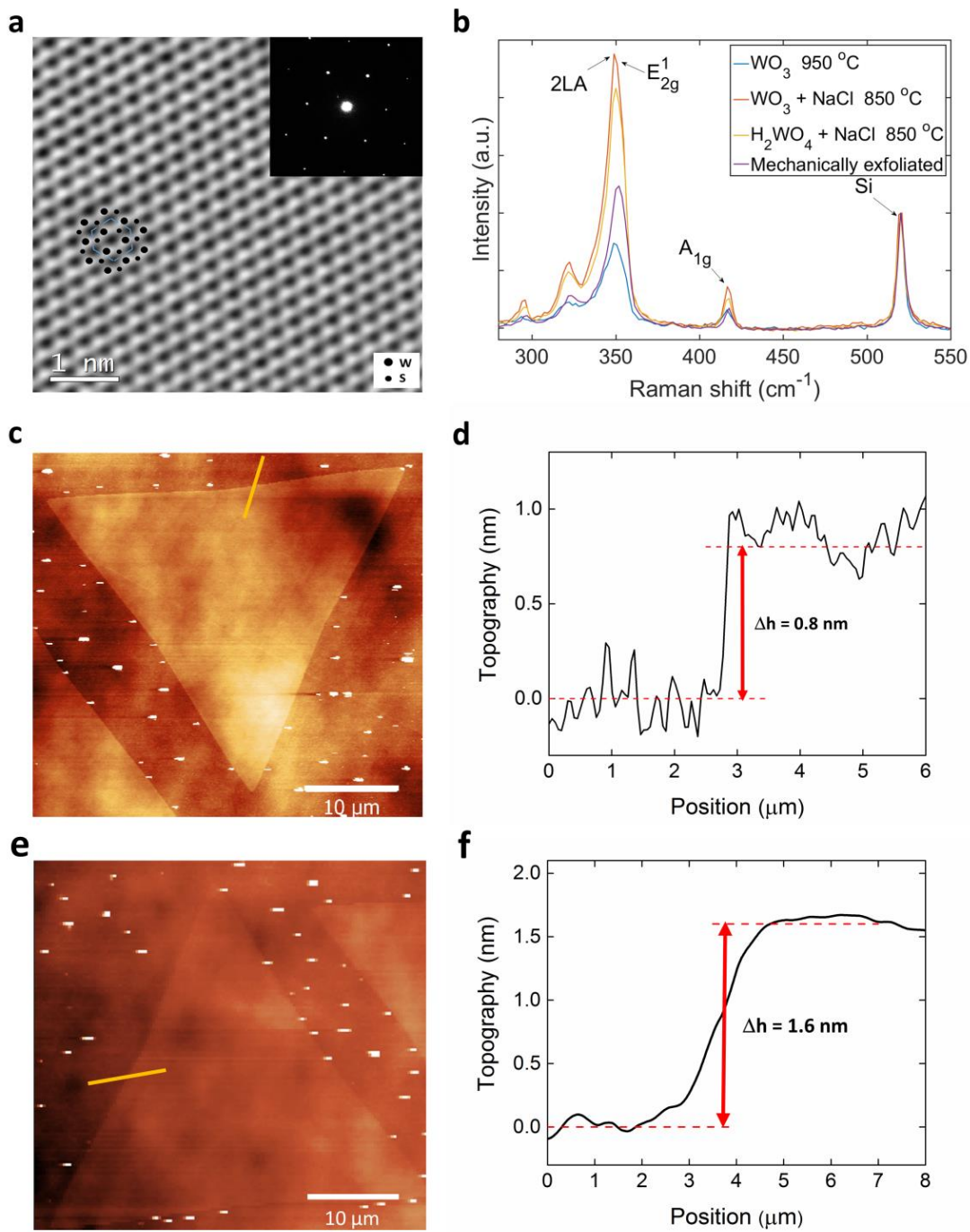

*Figure 2.* Structural and physical characterization of WS$_2$ triangles: (a) Phase image of the reconstructed exit-plane wave function from a focal-series of HRTEM images of the WS$_2$ lattice grown using H$_2$WO$_4$+NaCl, the inset reports a selected diffraction area which shows an hexagonal pattern; (b) Raman spectra showing the characteristics active modes of WS$_2$ grown under different

*conditions and compared with mechanically exfoliated flakes; (c) AFM image and (d) corresponding thickness profile of monolayer WS$_2$; (e) AFM image and (f) corresponding thickness profile of bilayer WS$_2$.*

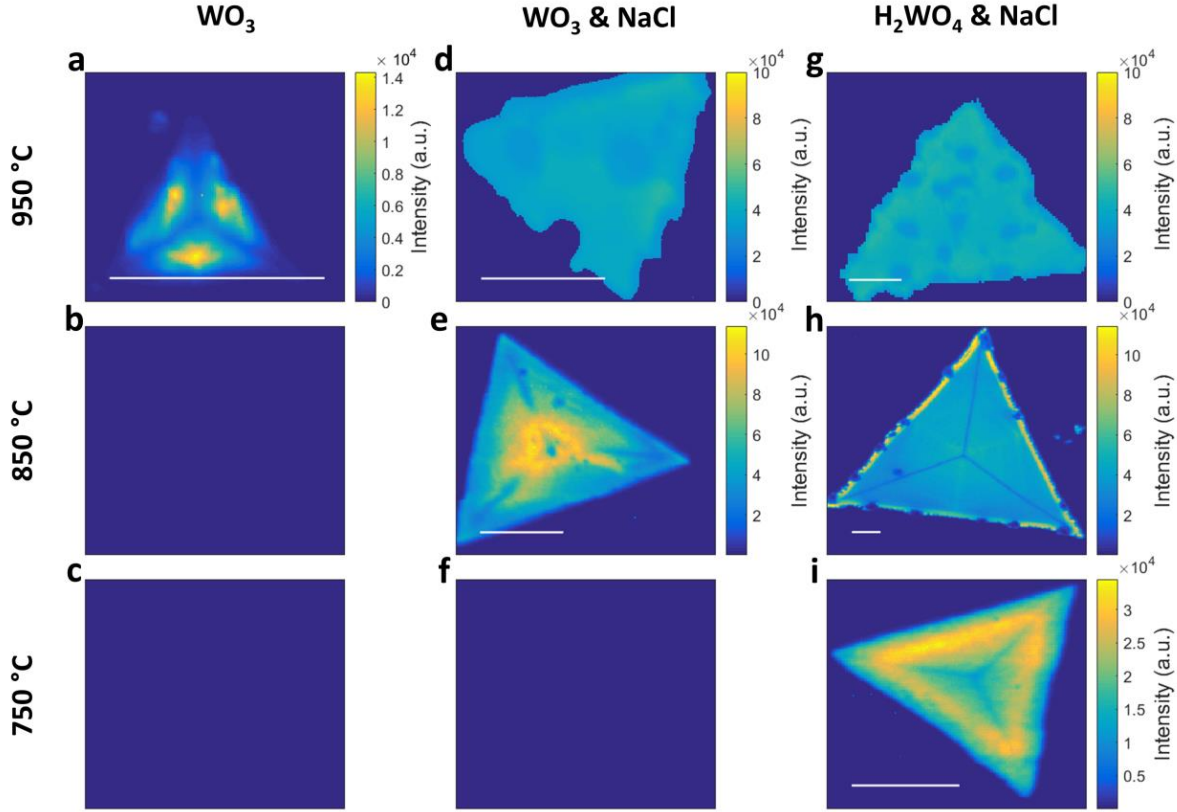

*Figure 3. Spatial maps of PL intensity of WS$_2$ grown in the conditions exemplified in Figure 1. The scale bar length is 10 μm.*

A comparison of representative photoluminescence (PL) intensity maps of WS$_2$ monolayers grown under the three precursors systems at different temperatures is reported in Figure 3. The PL peak intensity appears consistently higher for the NaCl-based precursor systems as compared to WO$_3$. The intensity variation pattern across an individual flake is not yet fully understood[41], and it is likely to be due different defect concentrations in the form of sulphur vacancies. The FWHM of the PL peaks significantly decreases from ~ 75 meV to ~ 50 meV to ~ 36 meV for the three precursors systems, $WO_3$, $WO_3$+NaCl, $H_2WO_4$ +NaCl respectively. They distribution is significantly narrow and uniform across the same flake and different flakes (SI). It is worth noting that 36 meV of FHWM is narrower that mechanically exfoliated material (Figure 4a) which is ~59 meV. This suggests that $WS_2$ grown by using $H_2WO_4$+NaCl possesses less structural defects compared to the other precursor systems, and specifically in the form of sulfur vacancies, which lead to a negligible contribution from trions to the PL peak. The PL peak position progressively blueshifts from 1.94 eV to 1.96 to 1.98 eV, reaffirming a decrease amount of S vacancies and the formation of a progressively more pristine material.

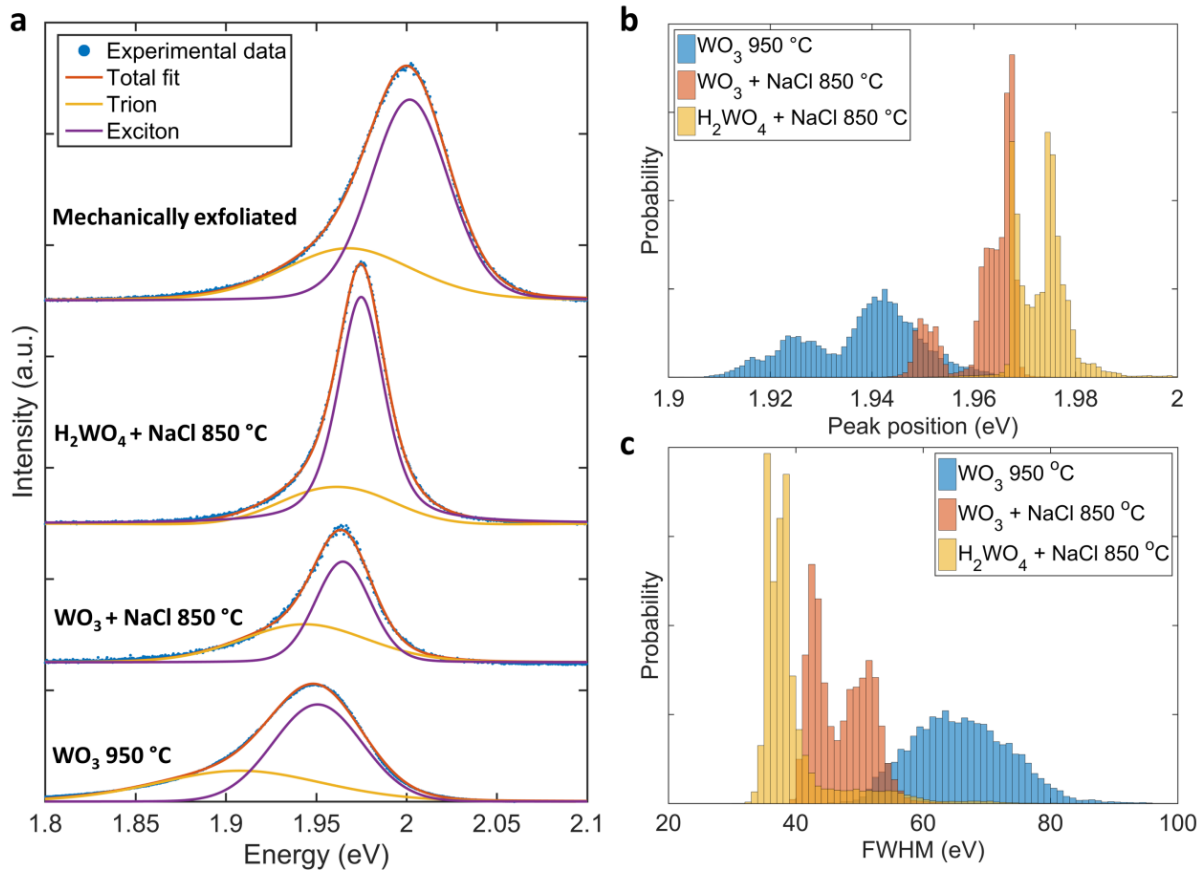

*Figure 4. PL spectra characteristics of $WS_2$ grown using: $WO_3$ at 950 °C, $WO_3$+NaCl at 850 °C, $H_2WO_4$+NaCl at 850 °C: (a) individual spectra (dotted line) and deconvolution in exciton and*

*trion components; (b) distribution of PL peak position and (c) distribution of PL FWHM for several WS$_2$ monolayers grown using the three different precursor systems.*

A molecular conversion based-growth mechanism, were tungsten oxyhalide molecules are sulfidized in vapour phase, versus a topotactic conversion of WO$_3$ in WS$_2$ can explain the different defects contents in WS$_2$. The greater efficiency of H$_2$WO$_4$ in inducing a complete sulfidization of the precursors compared with the WO$_3$+NaCl system has been also confirmed by chemical analysis (X-ray photoelectron spectroscopy). Analysing WS$_2$ grown using H$_2$WO$_4$+NaCl, the W 4f$_{5/2}$ and W 4f$_{7/2}$ core levels (Figure 5a) present peak position characteristic of W$^{4+}$ in WS$_2$[42,43] (32.7 and 34.8 eV respectively) and the narrowest achievable FWHM (1 eV) (Figure 5a), using the Mg K$\alpha$ as X-ray source. This indicates chemical purity and perfect stoichiometric ratio of W and S. This has been also confirmed by calculating the concentration of S and W from the integrated intensity of the W 4f and S 2p core levels. The S 2p$_{1/2}$ and 2p$_{3/2}$ core levels, also appear at the expected position for WS$_2$ (162.3 eV and 163.4 eV respectively, Figure 5b)[43] and with a very narrow FWHM (1 eV) (Figure 5c). A very small amount of W$^{6+}$ (W 4f$_{5/2}$ and W 4f$_{7/2}$ core levels centred at 35.9 eV and 38.1 eV respectively in Figure 5a) attributable to WO$_3$, which partially overlaps with the W 5p core level (38.5 eV), can be observed which however disappears after transferring the flakes on a new SiO$_2$/Si substrate (Figure 5c) and thus suggesting that it is related to residual precursors on the substrate considering the XPS spot size is ~1 mm. It is worth noting that after transfer the FWHM of the W 4f core levels remains unchanged suggesting that the transfer process preserves the crystallinity of the flakes and no additional defects are introduced.

Similarly, chemical purity and expected stoichiometric ratio of 2:1 for S:W have been observed for WS$_2$ grown from WO$_3$+NaCl (Figure 5a). Nevertheless, a larger W$^{6+}$ contribution, attributable to WO$_3$ (W 4f$_{5/2}$ and W 4f$_{7/2}$ centred at 35.9 eV and 38.1 eV respectively in Figure 5a) has been detected in this case suggesting that a conspicuous amount of precursors does not get sulfurized and it is just deposited onto the $SiO_2$ wafer. Upon transfer on a new $SiO_2/Si$ substrate, this component entirely disappears (Figure 5c), thus indicating also in this case that $WO_3$ is mainly distributed on the substrate. The FWHM of the $W^{4+}$ 4f core levels is ~1.2 eV in this case, suggesting higher concentration of defects compared to $H_2WO_4$+NaCl-led growth (Figure 5a). The transferred $WS_2$ present a FWHM even larger ~1.3 eV, suggesting the introduction of atomic defects as a consequence of the mechanical stress underwent by the flakes (Figure 5c). To conclude, XPS study confirms the effectiveness of $H_2WO_4$ as precursor versus $WO_3$.

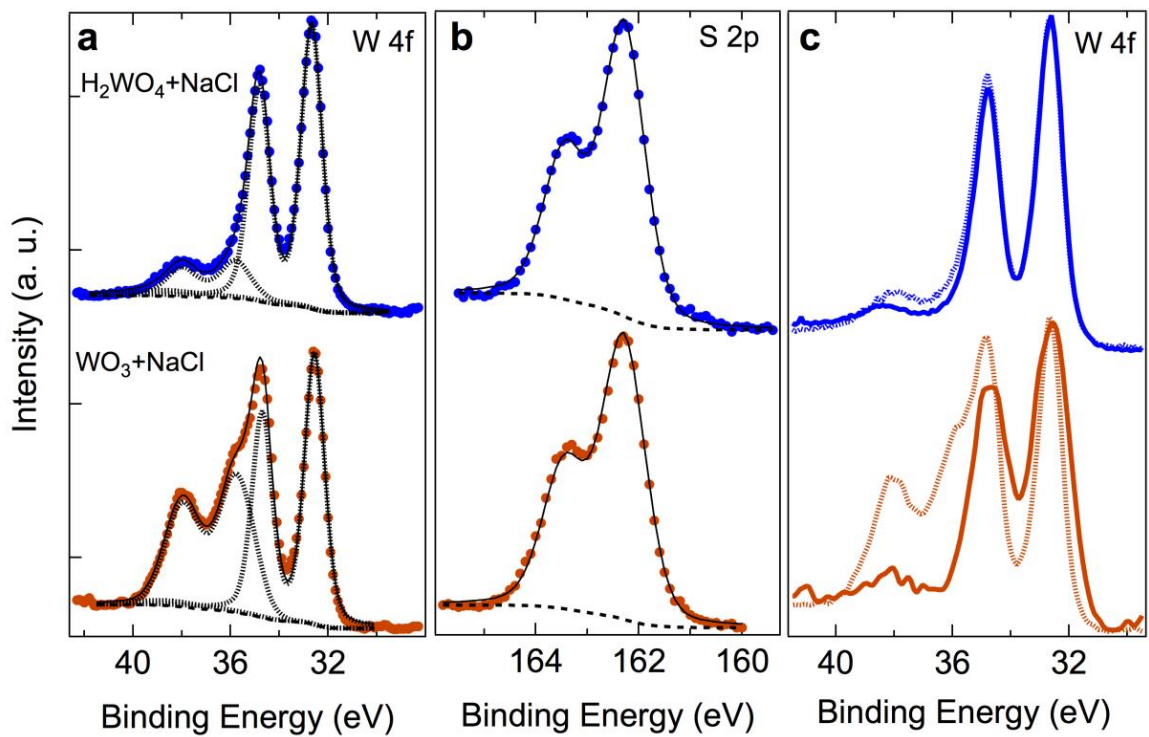

*Figure 5.* XPS spectra of the W 4f and S 2p core level peak regions. (a) Comparison of W $4f_{5/2}$, W $4f_{7/2}$ and W 5p core levels of $WS_2$ grown using $H_2WO_4$+NaCl at 950 °C (blue spectrum) with $WS_2$ grown using $WO_3$+NaCl at 950 °C (red spectrum). The deconvolution of W $4f_{5/2}$, W $4f_{7/2}$ and

*W 5p core levels and overall fit of the spectrum are reported as black dashed and a continuous line respectively. (b) The S $2p_{1/2}$ and $2p_{3/2}$ core levels for each of the two growth conditions are reported in the central panel. (c) W $4f_{5/2}$, W $4f_{7/2}$ and W 5p core levels before (dashed line) and after transfer (continuous like) onto a new $SiO_2$/Si substrate are compared showing the complete disappearance of the residual $WO_3$ components. The spectra were fit by Doniach-Sunjic function after subtracting a Shirley background (black dashed line).*

The progressive reduction of structural defects from using $WO_3$ to $H_2WO_4$ has been proven by electrical characterization. The electrical properties of the $WS_2$ flakes were characterised through their performance in bottom-gated field effect transistors (FET) (Figure 6a,b). The FET transfer curve (Figure 6c) displays an accumulation-type *n*-channel transistor, where the current flowing through the channel increases with increased gate bias, after the threshold voltage.

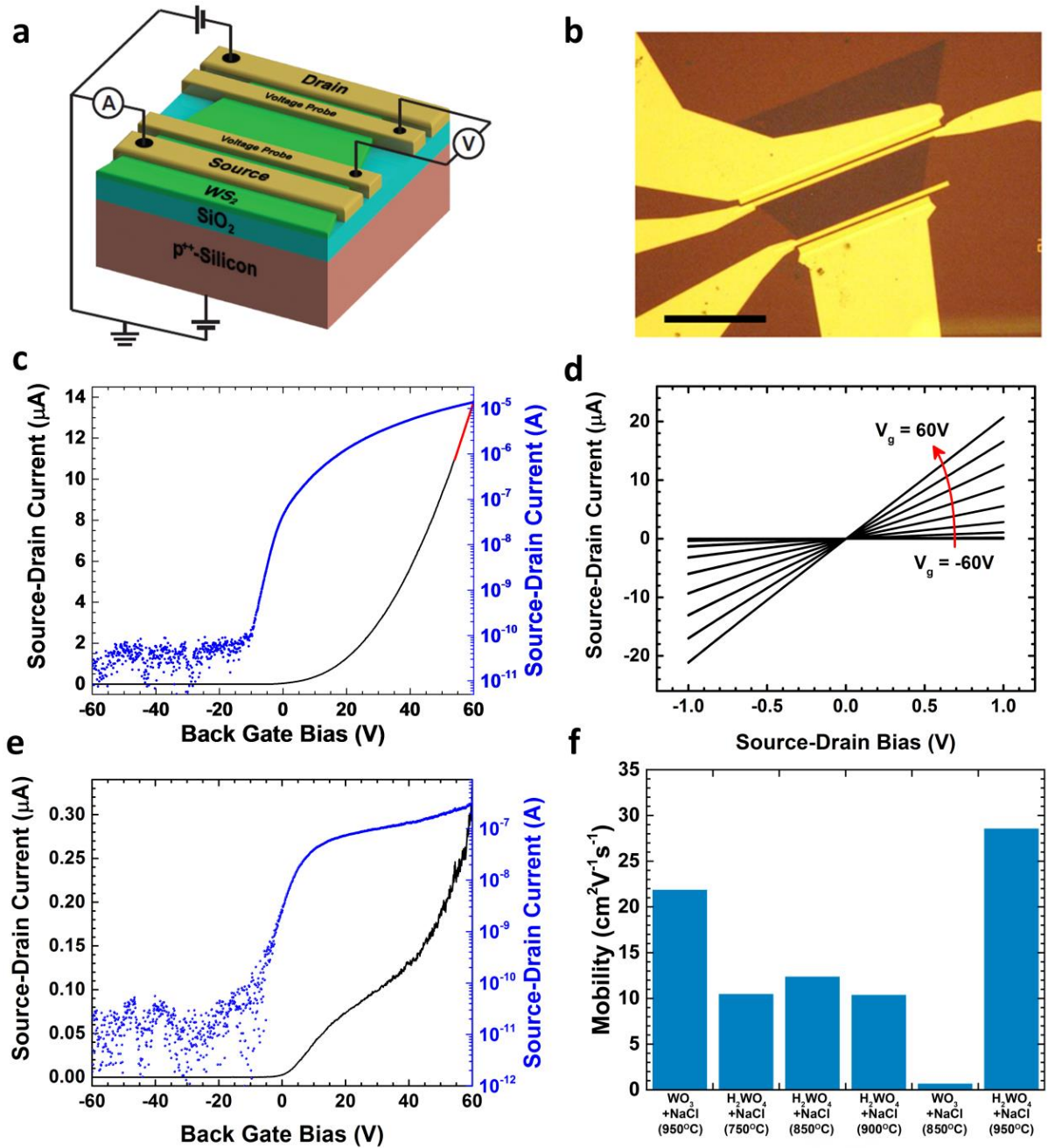

*Figure 6.* *Electrical characteristics of monolayer WS$_2$: (a) Schematic of the bottom-gated field effect transistors; (b) optical micrograph of the device (scale bar is 20 μm); (c) FET transfer curve for the monolayer WS$_2$ grown using H$_2$WO$_4$+NaCl at 950 °C showing the highest mobility of 28 cm$^2$/Vs (linear region of the transport graph marked with a red-dashed line); (d) Response curves at different gate biases for a WS$_2$ triangle grown using H$_2$WO$_4$+NaCl; (e) FET transfer curve for*

*the monolayer WS$_2$ grown using WO$_3$+NaCl at 850 °C; (f) electron mobilities of monolayers WS$_2$ grown using different conditions.*

The field-effect mobility was calculated in the linear region of the transport graph (marked with red-dashed line in Figure 6c), using $\mu_n = C_{ox}^{-1}(d\sigma/dV_{gs})$. Overall, monolayer WS$_2$ grown using H$_2$WO$_4$+NaCl shows electron mobilities systematically higher compared with the WO$_3$+NaCl system (Figure 6c,e,f) corroborating the fact that higher crystal quality is expected by using H$_2$WO$_4$ as precursor. Further, monolayer WS$_2$ presents electron mobility of 28 cm$^2$/Vs (Figure 6c,f) which is the highest mobility reported so far for CVD grown WS$_2$ deposited onto SiO$_2$ (Figure 7a)[19,21,23-25,30,44-48] and comparable to mechanically exfoliated WS$_2$[37,49-51]. The highest mobilities using either H$_2$WO$_4$+NaCl or WO$_3$+NaCl are displayed at 950 °C (Figure 6f) suggesting that the growth temperature does also play a role in improving the crystal quality of the material. While the role played by the different precursors systems in determining the crystallinity of the synthesis product becomes more prominent at low growth temperatures. Monolayer WS$_2$ grown using H$_2$WO$_4$+NaCl exhibits electron mobilities comprised between 10 and 20 cm$^2$/Vs at temperatures between 750 °C and 850 °C. While the electron mobilities of monolayer WS$_2$ grown using WO$_3$+NaCl at 850 °C (Figure 6f) present lower values ~2 cm$^2$/Vs. Bilayer WS$_2$ shows electron mobility systematically higher than monolayer (between ~38 cm$^2$/Vs and 52 cm$^2$/Vs) and also systematically higher than mechanically exfoliated bilayered flakes[51,52] (Figure 7b). The electron mobility of 52 cm$^2$/Vs (Figure 7b and Figure 8a,b,c) represents a record mobility as compared to CVD grown or mechanically exfoliated bilayer WS$_2$ onto SiO$_2$ reported so far[51-53]. The fact that the highest mobility for bilayer WS$_2$ has been obtained using WO$_3$+NaCl with no use of H$_2$WO$_4$ suggests that the bilayer system is less affected by the precursos choice, and a bilayered material presents in general crystal quality superior to monolayers.

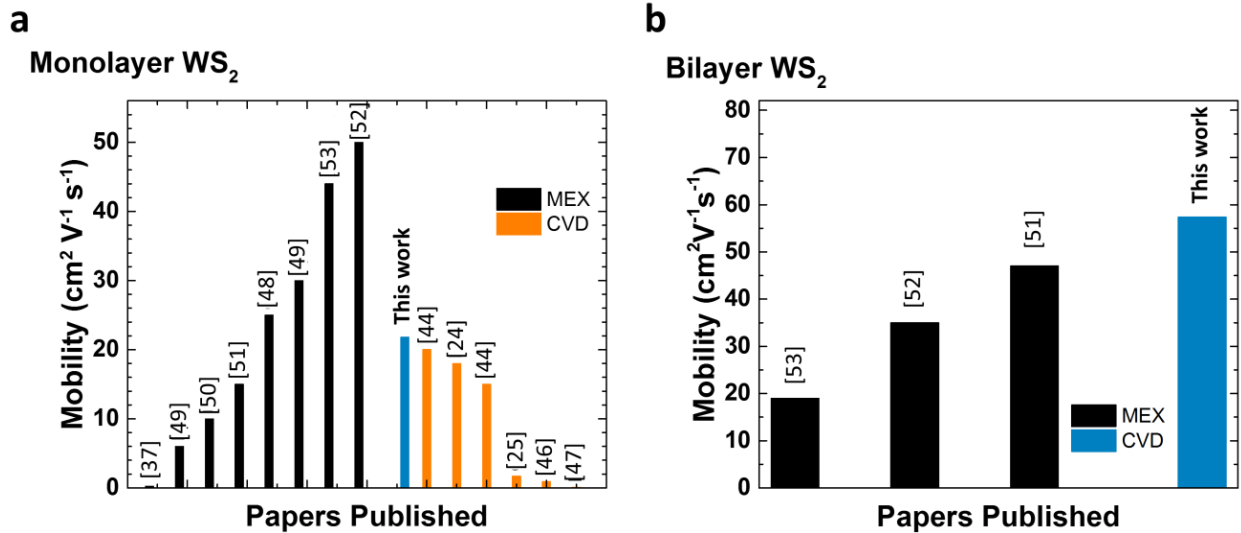

*Figure 7.* Comparison of our results with the literature of CVD grown material and mechanically exfoliated $WS_2$ (MEX): electron mobility for (a) monolayer $WS_2$ and (b) bilayer $WS_2$. The histograms show our record values for both monolayer and bilayer amongst the best values reported for CVD grown $WS_2$.

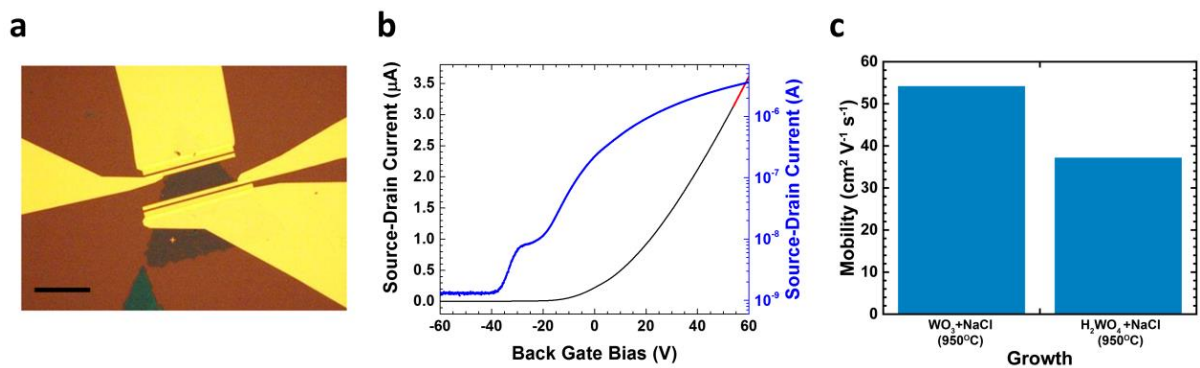

*Figure 8.* Electrical characteristics of bilayer $WS_2$: (a) Optical micrograph of the device (scale bar is 30 μm); (b) FET transfer curve for the bilayer $WS_2$ grown using $WO_3$+NaCl at 950 °C showing the highest mobility of 52 $cm^2$/Vs (linear region of the transport graph marked with a red-dashed line); (c) electron mobility of bilayer $WS_2$ grown by using different precursor systems.

## CONCLUSIONS

In conclusion, we have developed a synthesis strategy, which enables high crystal quality $WS_2$ as reflected in the high optical quality and in the carrier mobility that overcome naturally occurring materials. The molecular precursors approach leads to effective sulfidization of W, revealing to be highly advantageous with respect to the traditional oxide–based conversion synthesis of $WS_2$. These results can be translated and applied to the synthesis of different TMDs, and pave the way towards industrially scalable synthesis of monolayer $WS_2$ over large areas.

## METHODS

**CVD Synthesis of $WS_2$.** Commercial $WO_3$ (0.3 g, 99.9 %, Sigma Aldrich), $H_2WO_4$ (0.3 g, 99.9 %, Sigma Aldrich) and NaCl (0.3 g, ≥99.5 %, Sigma Aldrich) powders were loaded in an alumina boat placed in the center of a 2 inch-diameter horizontal quartz tube CVD furnace. While an alumina boat containing sulfur powders (0.6 g, ≥99.5 %, Sigma Aldrich) was loaded in the upstream zone of the tube, whose temperature was independently controlled by a different heater. The growth substrates were Si wafers (500 microns) with on top 285 nm of $SiO_2$ thermally deposited. The substrates were sequentially cleaned for 15 min in acetone, isopropanol and deionized water in a sonicator, followed by dipping in $H_2SO_4/H_2O_2$ (3:1) for two hours and rising in deionized water. They were then blow dried with nitrogen gas, cleaned with $O_2$ plasma for 5 min and loaded into the downstream zone of the furnace. The CVD growths were then performed at low pressure (~$10^{-1}$ mbar) and under flow of high purity Ar gas (flow rate of 100 sccm). The furnace was heated to 750-950 °C with a ramp rate of 25 °C min$^{-1}$, kept at the growth temperatures for 15 min and then naturally cooled down to room temperature. The sulfur powder was independently heated to 125 °C with a ramp of 5 °C min$^{-1}$, kept at this temperature for 15 min and naturally cooled down.

**Sample Transfer.** The transfer procedure was performed by depositing a PMMA film 350 nm thick onto the target sample, which was subsequently immersed in a KOH solution (0.1M) until detachment of the PMMA from the SiO$_2$/Si substrate. The PMMA/WS$_2$ films was then scooped out with a new Si/SiO$_2$ substrate, repeatedly washed in deionized water and then immerged in an acetone bath at 45 °C for 20 min to dissolve the PMMA film.

**TEM Characterization.** TEM analysis of the WS$_2$ flakes was carried out on a FEI Titan 80-300 S/TEM operated at 80 kV, equipped with a monochromator and a Cs aberration image corrector. Focal series micrographs of the representative flake were acquired at different objective lens focus values (using a spherical aberration coefficient Cs ~ -4 µm) and exit-wave reconstruction was performed using TrueImage software (FEI).

**Physical Characterization.** Raman and photoluminescence spectra were collected using a Renishaw inVia spectrometer equipped with a 532 nm laser excitation. All the spatial maps were collected under a 100x objective using grating of 1800 line/mm, which provide a resolution of ~1.5 cm$^{-1}$.

**X-ray Photoelectron Spectroscopy.** X-Ray photoemission spectra were acquired in a custom made ultra-high vacuum system (pressure < 10$^{-9}$ mbar) equipped with a VG Escalab Mk-II electron analyzer and a twin anode (Al/Mg) non-monochromatized x-ray source (Omicron DAR 400). All measurements were taken in quasi-normal emission (5° off) at room temperature using a pass-energy of 20 eV, an energy step of 0.1 eV and the Mg K$_\alpha$ emission line as exciting radiation.

**Device Fabrication.** For the field-effect mobility measurements, single and bilayer WS$_2$ field-effect transistors (FET) were fabricated after transfer to a new Si/SiO$_2$ substrate where the Si is highly *p*-doped and acts as a global gate electrode, and a 285 nm thick thermally grown SiO$_2$ serves as the gate dielectric. The FETs were fabricated on dried Si/SiO$_2$/WS$_2$ samples. Besides the current bearing contacts, termed here "Source" and "Drain", two additional voltage probes were added to each FETs to allow for an accurate determination of the channel's conductance by eliminating the contribution of the contacts. Both the current bearing leads and the voltage probes were patterned simultaneously using a standard electron beam lithography process. The source, drain and voltage probes consisted of 50 nm Au, while the electronics leads consisted of 5 nm Ti and 50 nm Au. A two-steps annealing process followed the fabrication. The samples were first annealed for 2 hours at 200 °C under H$_2$/Ar (10/90) flow in atmospheric pressure, to etch residues of PMMA that was used as an electron resist for the lithography step. Then, the samples were put under a high vacuum (~$10^{-6}$ mbar) and baked at 115 °C for 60 hours, to promote desorption of water molecules from the channel surface.

**Electrical Measurements.** The FETs were measured inside the vacuum chamber at a constant pressure of ~$10^{-6}$ mbar, without exposure to ambient conditions after the second annealing step. The drain electrode was biased with a low noise source-meter and the source electrode was grounded throughout the experiment. An additional source-meter was used to bias the global gate electrode, with respect to the source. The transistor current, $I_{ds}$, was measured using an ammeter and the potential difference across the voltage probes, $V_{A-B}$, was measured with a voltmeter. The channel conductivity, σ, is then readily obtained using σ = ($L\ I_{ds}$)/($W\ V_{A-B}$), where $W$ and $L$ are the channel's width and length, respectively. The measurement set-up is shown schematically (not to scale) in Figure 6a. The oxide capacitance was estimated to be 115 μFm$^{-2}$ from $C_{ox} = \varepsilon_0\ \varepsilon_r/d_{ox}$, where $d_{ox}$ is the oxide thickness and $\varepsilon_0$ and $\varepsilon_r$ are the vacuum permittivity and $SiO_2$ relative permittivity, respectively.

## ACKNOWLEDGMENTS

C.M. would like to acknowledge the EPSRC awards EP/K033840/1, EP/K01658X/1, EP/K016792/1, EP/M022250/1 and the EPSRC-Royal Society Fellowship Engagement Grant EP/L003481/1. C.M. acknowledges the award of a Royal Society University Research Fellowship by the UK Royal Society. N.N. acknowledge the Imperial College Junior Research Fellowship and P.C.S. would like to acknowledge the funding and support from the European Commission (H2020 – Marie Sklodowska Curie European Fellowship - 660721). I.A. acknowledges financial support from The European Commission Marie Curie Individual Fellowships (Grant number 701704). J.D.M. acknowledges the financial support from the Engineering and Physical Sciences Research Council (EPSRC) of the United Kingdom, via the EPSRC Centre for Doctoral Training in Metamaterials (Grant No. EP/L015331/1). S.R. and M.F.C acknowledge financial support from EPSRC (Grant no. EP/J000396/1, EP/K017160/1, P/K010050/1, EP/G036101/1, EP/M001024/1, EP/M002438/1), from Royal Society international Exchanges scheme 2016/R1 and from The Leverhulme trust (grant title "Quantum Drums" and "Room temperature quantum electronics").